\documentstyle[emulateapj,apjfonts,psfig]{article}

\lefthead{J.~M.~Miller et al.}  \righthead{Cool Disks in ULX Sources}

\received{Date}
\revised{Date}
\accepted{Date}

\journalid{vol}{date}
\articleid{1}{4}
\paperid{id}

\cpright{AAS}{1999}
\ccc{x}

\begin{document}

\title{X-ray Spectroscopic Evidence for Intermediate Mass Black Holes:\\
Cool Accretion Disks in Two Ultra--Luminous X-ray Sources}

\author{J.~M.~Miller\altaffilmark{1,4}, 
        G.~Fabbiano\altaffilmark{1},     
	M.~C.~Miller\altaffilmark{2},
        A.~C.~Fabian\altaffilmark{3}
        }

\altaffiltext{1}{Harvard-Smithsonian Center for Astrophysics, 60
        Garden Street, Cambridge, MA 02138, jmmiller@head-cfa.harvard.edu}
\altaffiltext{2}{Department of Astronomy, University of Maryland,
        College Park, MD, 20742}
\altaffiltext{3}{Institute of Astronomy, University of Cambridge,
        Madingley Road, Cambridge CB3 OHA, United Kingdom}
\altaffiltext{4}{National Science Foundation Astronomy and
        Astrophysics Fellow}

\keywords{Black hole physics -- X-rays:stars}

\authoremail{jmmiller@head-cfa.harvard.edu}

\label{firstpage}

\begin{abstract}
We have analyzed an {\it XMM-Newton} observation of the nearby spiral
galaxy NGC 1313, which contains two ``ultra-luminous'' X-ray (ULX)
sources.  We measure isotropic luminosities of $L_{X} = 2.0 \times
10^{40}~erg~s^{-1}$ and $L_{X} = 6.6 \times 10^{39}~erg~s^{-1}$ for
NGC 1313 X-1 and X-2 (0.2-10.0 keV, assuming a distance of 3.7~Mpc).
The spectra {\it statistically require} soft and hard spectral
components to describe the continuum emission; some prior studies of
ULXs have claimed cool soft components with lower statistics.  The
improvement over several single-component models exceeds the 8$\sigma$
level of confidence for X-1; the improvement for X-2 is significant at
the 3$\sigma$ level.  The soft components in these ULX spectra are
well-fit by multi-color disk blackbody models with color temperatures
of $kT \simeq 150$~eV.  This temperature differs markedly from those
commonly measured in the spectra of stellar-mass ($10~M_{\odot}$)
black holes in their brightest states ($kT \simeq 1$~keV).  It is
expected that the temperature of an accretion disk orbiting a black
hole should decrease with increasing black hole mass.  If the soft
components we measure are due to emission from the inner region of an
accretion disk, and disks extend close to the innermost stable
circular orbit at the accretion rates being probed, the low color
temperatures may be interpreted as spectroscopic evidence of black
holes with intermediate masses: $M_{BH} \simeq 10^{3}~M_{\odot}$.
Simple Eddington scaling arguments suggest a mimum mass of $M_{BH}\sim
10^2 M_{\odot}$.  NGC 1313 X-1 and X-2 are found in optical nebulae
(Pakull \& Mirioni 2002), which may indicate that anisotropic emission
geometries are unlikely to account for the fluxes observed.
\end{abstract}

\section{Introduction}
ULX sources may be defined as point-like off-nuclear X-ray sources in
normal galaxies for which measured luminosities exceed the isotropic
Eddington limit for a stellar-mass (less than approximately
$10~M_{\odot}$) black hole.  The existence of such sources was first
revealed with {\it Einstein} (Fabbiano 1989).  Variability has been
observed in many ULX sources on the timescales of months and years (in
rare cases, on shorter timescales), which suggests that they are
accreting objects.

Although black holes with intermediate mass (IMBHs;
$10^{2-5}~M_{\odot}$) provide an attractive explanation for the nature
of ULX sources, strong evidence for this interpretation has been
lacking --- especially from X-ray spectroscopic studies.  Anisotropic
emission from stellar-mass black holes may be able to account for the
flux observed in some ULX sources (King et al. 2001).
Single-component fits with the multi-color disk blackbody model (MCD;
Mitsuda et al. 1984) have measured color temperatures above those
commonly reported in stellar-mass black holes ($kT = 1-2$~keV; see,
e.g., Sobczak et al. 2000, Makishima et al. 2000); as inner disk
temperatures should fall with increasing black hole mass, these
findings again point towards stellar-mass black holes or Kerr black
holes.

NGC 1313 is a nearby spiral galaxy ($d = 3.7$~Mpc, Tully 1988).  Two
ULX sources --- NGC 1313 X-1 X-2 --- are associated with this galaxy;
approximately 1~kpc and 8~kpc from the photometric center of the
galaxy, respectively (Colbert et al. 1995).  Variability was observed
in {\it ROSAT}/HRI observations of these sources (Colbert \& Ptak
2002).  Colbert and Mushotzky (1999) claim evidence for a cool ($kT =
120$~eV) disk in fits to an {\it ASCA} spectrum of NGC 1313 X-1 with a
model consisting of MCD and power-law components; however the spectrum
is fit acceptably by a simple power-law ($\Gamma = 1.74$,
$\chi^{2}/{\rm d.o.f.} = 244/259$).  Similarly, a cool disk may be
implied in joint fits to {\it BeppoSAX}/MECS and {\it ROSAT}/PSPC
spectra of M81 X-9 (La Parola et al. 2001) but not in fits to MECS and
LECS spectra (both aboard {\it BeppoSAX}).

We have analyzed the archival {\it XMM-Newton} spectra of NGC 1313 X-1
and X-2.  Separate soft and hard components are statistically required
to describe the spectra.  We measure low temperatures ($kT \simeq
150$~eV) for the soft components.  We interpret the soft component as
arising from the inner region of an accretion disk, and explore the
implications of cool accretion disks for the mass of black holes, if
such objects power X-1 and X-2.

\section{Data Reduction and Analysis}
NGC 1313 was observed by {\it XMM-Newton} on 17 October 2000 starting
at 03:59:23 (UT).  We used only the EPIC data for this analysis.  The
EPIC cameras were operated in ``PrimeFullWindow'' mode with the
``medium'' optical blocking filter.  The {\it XMM-Newton} reduction
and analysis suite SAS version 5.3.3 was used to filter the standard
pipeline event lists, to detect sources within the field, and to make
spectra and responses.  The pipeline processing and our own both
failed to produce an event list for the pn camera.  We therefore
restricted our analysis to the MOS-1 and MOS-2 cameras.  Application
of the standard time filtering resulted in a net exposure of
29.3~ksec.

The source locations were determined by running the SAS tool
``edetect\_chain''.  With this tool, we find NGC 1313 X-1 at
$3{h}18{m}19.99{s},~ -66^{\circ}29'10.97''$, and NGC 1313 X-2 at
$3{h}18{m}22.34{s},~ -66^{\circ}36'03.68''$ (J2000).  The tool returns
an error of 0.2'' for these positions; an uncertainty of 4'' may be
more appropriate (e.g., Foschini et al. 2002).  Source counts were
extracted in a circle within 24'' of the detected source position.
Background counts were extracted in an annulus between 24''--30''.  To
create spectra, we then applied the selection criteria described in
the MPE ``cookbook.''  These selections are as follows: we set
``FLAG=0'' to reject events from bad pixels and events too close to
the CCD chip edges, event patterns 0--12 were allowed, and the MOS
spectral channels were grouped by a factor of 15.  Response files were
made using the SAS tools ``rmfgen'' and ``arfgen.''  Spectral files
were grouped to require at least 20 counts per bin before fitting to
ensure the validity of $\chi^{2}$ statistics.

Model spectra were fit to the data using XSPEC version 11.2 (Arnaud
1996).  The MOS-1 and MOS-2 spectra were fit jointly with an overall
normalizing constant.  The constant indicates that the overall flux
normalizations of these cameras differ by less than 5\%.  Models were
fit to the spectra in the 0.2--10.0~keV band.  Systematic errors were
not added to the spectra.  Errors quoted in this work are at the 90\%
confidence level.  Using the SAS tool ``epatplot'' we found photon
pile-up to be negligible in our source spectra; this was confirmed
with the HEASARC tool ``PIMMS'' using the parameters reported by
Colbert \& Mushotzky (1999) for X-1 and X-2.

\section{Results}
Lightcurves of the source event lists do not show strong variability
on the timescale of this observation.  We therefore proceeded to make
fits to the time-averaged spectra.  The results of joint fits to the
MOS-1 and MOS-2 spectra in the 0.2--10.0~keV band are listed in Table
1.

We began by fitting single-component models commonly applied to ULX
sources, modified by photoelectric absorption (via the ``phabs'' model
within XSPEC).  For both X-1 and X-2, MCD, thermal Bremsstrahlung, and
Raymond-Smith plasma models all fail to yield acceptable fits.  Simple
power-law models provide improved but statistically unacceptable fits.
Broken power-law models with breaks in the 0.2--10.0~keV range also
yield improved but statistically unacceptable fits.

An observation of NGC 1313 with {\it Chandra} finds that the diffuse
emission in our {\it XMM-Newton} extraction region is not
well-described with any thermal model; the best model is a $\Gamma =
1.7$~ power-law with an unabsorbed flux of $F_{0.3-7.0} = 5.9 \times
10^{-14}$~erg/s (A. Kong, priv. comm.).  This signals that the soft
components may indeed arise in accretion disks, and we explored fits
with a model consisting of MCD and power-law components, and the
``CompTT'' model.  The MCD model describes a standard Shakura-Sunyaev
(1973) accretion disk as a series of blackbody annuli.  The latter model
describes Compton-upscattering in a corona of optical depth $\tau$
with electron temperature $kT_{corona}$ from a Wien distribution of
soft seed photons (Titarchuk 1994); we used the version of the model
which assumes a disk geometry for the seed photons.  The coronal
temperature could not be constrained and we therefore fixed
$kT_{corona} = 50$~keV, which is a moderate value.  X-2 is fit
acceptably by both of these models; the MCD plus power-law model is a
significantly better fit to the spectrum of NGC X-1.  The inability of
single-component models to describe the spectra 

\centerline{~\psfig{file=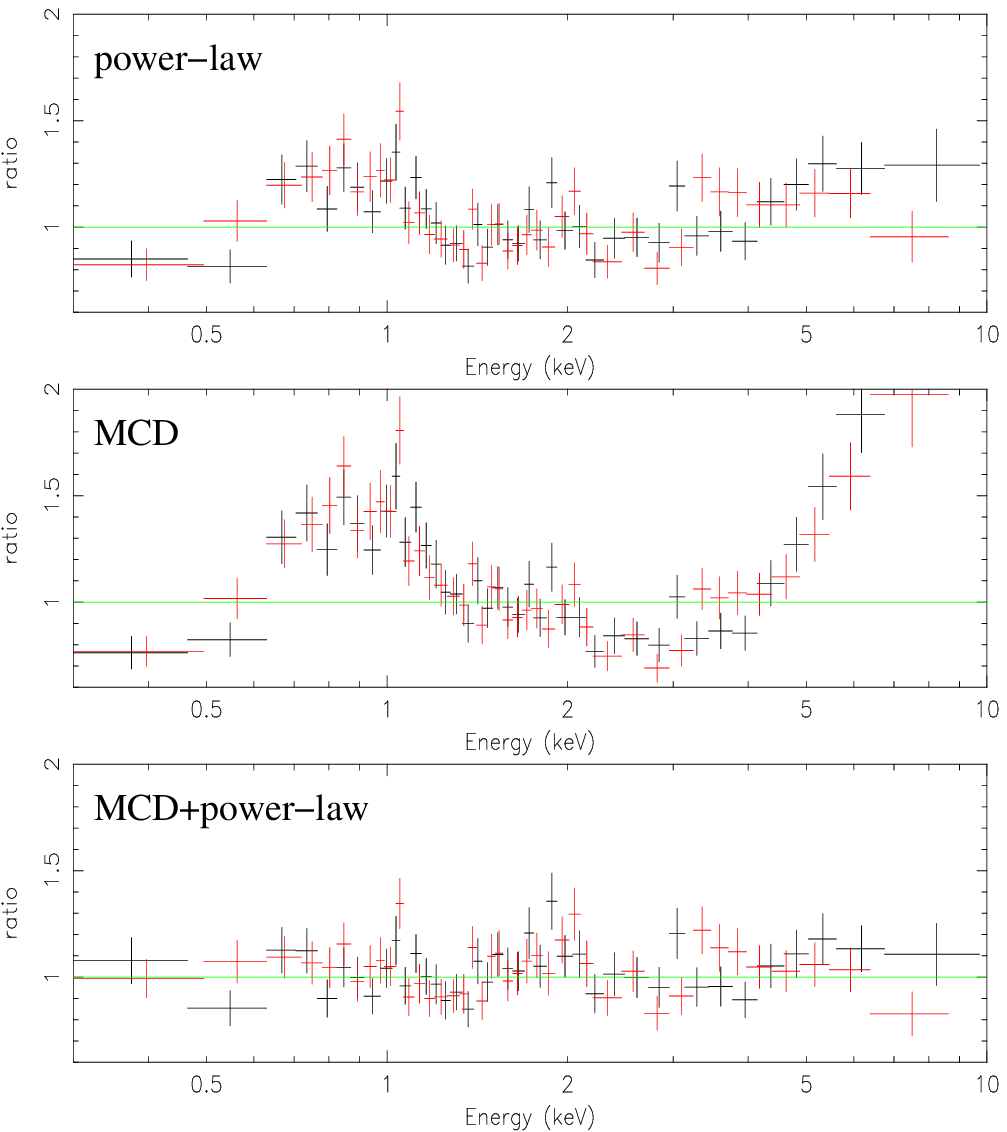,width=3.5in}~}
\figcaption[h]{\small The data/model ratios from fits to the MOS-1
(black) and MOS-2 (red) spectra of NGC 1313 X-1 with standard ULX
models (the spectra are rebinned for visual clarity).  The ratios
correspond to fits with: a simple power-law, the multicolor disk
blackbody model, and a model with both components (see Table 1).}
\medskip

of X-1 is demonstrated in Figure 1; MCD plus power-law fits to X-1 and
X-2 are shown in Figures 2 and 3.  The F-test indicates that the
improved fits given by the MCD plus power-law model are significant
over single-component models at more than the 8$\sigma$ level of
confidence for X-1, and at the 3$\sigma$ level of confidence for X-2.
With the MCD plus power-law model, we measure isotropic luminosities
of $L_{X-1} = 2.0^{+1.6}_{-0.3} \times 10^{40}$ erg/s and $L_{X-2} =
6.6^{+1.8}_{-2.0} \times 10^{39}$ erg/s (0.2--10.0 keV, assuming
$d=3.7$~Mpc).  If ULX sources are like Galactic black holes and AGN, a
corona may be the source of hard X-rays.  Extrapolating to the
0.05--100.0~keV band, we find that isotropic luminosities increase by
approximately 50\%.

We note that some HII regions within NGC 1313 may have low metallicity
(50--75\% relative to solar abundances; Zaritsky, Kennicutt, \& Huchra
1994).  The metallicity assumed for absorbing material can impact
low-energy spectral measurements.  For X-1, our best-fit cool disk
plus power-law model represents an improvement at the 5$\sigma$ level
of confidence relative to simple single-component fits with the
abundances of O and Fe fixed at $0.25 \times {\rm solar}$ in the
absorption model.  With such an absorption model, single-component
fits to the spectra of X-2 are acceptable.  The low-abundance
absorption model is likely inappropriate, however; fits with this
model under-estimate the depth of the O edge in the X-1 spectra.  The
Galactic column density along the line of sight to NGC 1313 is $4.0
\times 10^{20}~{\rm cm}^{-2}$; our significantly higher measurements
may indicate significant neutral absorption within NGC 1313 or
intrinsic to the sources (see Table 1).

The most remarkable result of our analysis is that both models require
cool accretion disks ($kT \simeq 150$~eV).  These temperatures differ
from those derived from the spectra of stellar-mass 

\centerline{~\psfig{file=f2.ps,width=3.5in,angle=-90}~}
\figcaption[t]{\small The unfolded MOS-1 and MOS-2 spectra of NGC 1313
X-1.  The total spectrum, cool ($kT \simeq 150$~eV) disk component,
and power-law components are shown in black, blue, and red,
respectively.}
\medskip

black holes in bright states; in such spectra $kT \simeq 1$~keV is
typical (e.g., Sobczak et al. 2000) but temperatures can reach to
nearly 2~keV in extreme cases.  These temperatures also differ
considerably from those measured with single-component fits with the
MCD model to the spectra of some ULXs, which approach $kT \simeq
2$~keV (e.g., Makishima et al. 2000).

The MCD model is based on the following relationship:
$T \propto M^{-1/4}$.  This can be used for scaling by writing
$(M_{ULX} / M_{10~M_{\odot}}) \propto (kT_{10~M_{\odot}} /
kT_{ULX})^{4}$.  If we assume that $kT = 1$~keV is a typical inner
disk color temperature for $10~M_{\odot}$ black holes for values of
$L_{X} / L_{Edd.}$ similar to those being observed in ULX sources, the
best-fit MCD color temperatures measured for X-1 and X-2 then imply
black hole primaries with masses near $M_{BH} \sim 2\times
10^{4}~M_{\sun}$.  If we assume $kT = 0.5$~keV may be more appropriate
for $10~M_{\odot}$ black holes at high accretion rates, this scaling
still suggests masses of $M_{BH} \sim 1.2 \times 10^{3}~M_{\sun}$ for
X-1 and X-2.

Shimura \& Takahara (1995) derived a hardening factor correction to
account for the effects of opacity on the measured disk temperature
and inner radius.  This correction is very simple: $kT_{corr.} =
f^{-1} kT_{obs.}$, and $R_{in, corr.} = \eta f^{2} R_{obs.}$~ (where
$f = 1.7$ is the hardening factor, and $\eta = 0.63$ is valid for $i <
70^{\circ}$ and accounts for the difference between the innermost
radius and the radius of peak temperature; see Sobczak et al. 2000 and
Makishima et al. 2000).  Merloni, Fabian, \& Ross (2000) have found
that hardening corrections may not be constant against changes in the
physical parameters of the disk.

The normalization of the MCD model allows for more direct estimates of
the black hole masses if we assume that $R_{in} = R_{ISCO} = 8.85~{\rm
km}~(M_{BH}/M_{\odot})$~(where $R_{ISCO}$ is the radius of the
innermost stable circular orbit), appropriate for Schwarzschild holes.
Simply, $M_{BH} = \eta f^{2} (K/{\rm cos}[i])^{1/2} \times (d/10~{\rm
kpc}) \times (8.85~{\rm km})^{-1}$, where $K$ is the model
normalization, and $i$ is the inclination of the system.  Using the
90\% confidence lower limit MCD normalizations (see Table 1) we find
lower-limit masses of $M_{X-1} \gtrsim 2200~M_{\odot}$ and $M_{X-2}
\gtrsim 830~M_{\odot}$ (assuming $i=0$ and $d=3.7$~Mpc; note that
higher values of $i$ and $d$ would {\it increase} the minimum mass
estimates, as would significant black hole spin).

Defining $L_{Edd.} = 1.3~(M_{BH}/M_{\odot}) \times 10^{38}$ erg/s
(Frank, King, \& Raine 2002) and taking $L_{X}$ in the 0.05--100.0~keV
range 

\centerline{~\psfig{file=f3.ps,width=3.5in,angle=-90}~}
\figcaption[t]{\small The unfolded MOS-1 and MOS-2 spectra of NGC 1313
X-2.  The total spectrum, cool ($kT \simeq 160$~eV) disk component,
and power-law components are shown in black, blue, and red,
respectively.}
\medskip

as a better approximation to a bolometric luminosity, we find
lower-limit isotropic luminosity masses of $M_{X-1} \gtrsim
230~M_{\odot}$ and $M_{X-2} \gtrsim 70~M_{\odot}$.  These mass
estimates are roughly an order of magnitude below those obtained by
scaling from the MCD fit parameters; it is possible that X-1 and X-2
are observed at $0.1~L_{Edd.}$, similar to many Galactic black holes.
We caution that the mass limits are only as good as the MCD model and
our best-fit continuum models.

\section{Discussion}
We have analyzed the EPIC MOS-1 and MOS-2 spectra of the ULX sources
NGC 1313 X-1 and X-2.  The spectra statistically require soft and hard
components to describe the continuum emission.  When the soft
components in X-1 and X-2 are fit with models for accretion disks, low
disk temperatures are obtained ($kT \simeq 150$~eV).  Scaling these
temperatures and the normalization of the MCD model suggests that X-1
and X-2 harbor black holes with $M_{BH} \simeq 10^{3}~M_{\odot}$, or
higher.  Isotropic Eddington luminosity scaling suggests that X-1 and
X-2 may harbor black holes with masses on the order of $M_{BH} \simeq
10^{2}~M_{\odot}$.  It is possible that we have observed X-1 and X-2
at $L_{X} \simeq 0.1 L_{Edd.}$; however, it is not clear which scaling
method is superior.

Colbert et al. (1995) estimate the radio power at the position of X-1
to be $10^{19}$~W/Hz at 1.4~GHz, implying an isotropic radio
luminosity of $10^{35}$ erg/s.  The radio to X-ray luminosity ratio is
then approximately $5 \times 10^{-6}$; this makes it unlikely that
relativistic beaming can account for the flux of X-1, because beaming
tends to produce flat $\nu F_{\nu}$ spectra (Fossati et al. 1998).
Pakull and Mirioni (2002) have found that X-1 lies in the center of a
diffuse H$_{\alpha}$ nebula with a radius of approximately 240~pc, and
a high [O~I] $\lambda$6300/H$_{\alpha}$ ratio implying X-ray
photoionization.  An optical survey of massive young star clusters in
nearby galaxies by Larsen (1999) reveals no cluster candidates within
approximately 380~pc of our position for X-1.  Similarly, Pakull \&
Mirioni (2002) find that X-2 lies at the center of an H$_{\alpha}$
nebula with strong [Si~II] and [O~I] lines, implying that X-2 is
acting on the local interstellar medium.  Larsen (1999) reports no
clusters within a few kpc of X-2.  These findings suggest that X-1 and
X-2 may emit nearly isotropically and illuminate their local nebulae.
It is unlikely, then, that models based on anisotropic emission from
stellar-mass black holes (e.g., King et al. 2001) can explain the
observed fluxes.

IMBHs may be the endpoints of very massive low-metallicity stars
(Heger et al. 2002), or perhaps Population III stars from the era of
galaxy formation (Madau \& Rees 2001).  Miller and Hamilton (2002)
have suggested that IMBHs may grow in globular clusters.  Ebisuzaki et
al. (2001) have proposed that intermediate mass black holes may form
in young compact star clusters.  If the disk temperature and
normalization scalings are correct, our estimates for the masses of
X-1 and X-2 are inconsistent with black holes in X-1 and X-2 being the
endpoints of low-metallicity or Population III stars; X-1 and X-2 may
be the result of growth by mergers.

Though a power-law produced a statistically acceptable fit to the {\it
ASCA} spectrum of X-1 ($\chi^{2}/{\rm d.o.f.} = 244/259$), the values
we have measured with the MCD plus power-law model are broadly
consistent with those reported by Colbert \& Mushotzky (1999) using
the same model.  We measure a lower disk temperature and harder
power-law index in X-2.  Makishima et al. (2000) fit two {\it ASCA}
spectra of NGC 1313 X-2 separated by two years with the MCD model, and
found $kT = 1.47$~keV and $kT = 1.07$~keV.  Our fits to the {\it
XMM-Newton} spectra of X-2 with only an MCD component are not
acceptable, but we measure $kT = 0.91$~keV --- consistent with the
latter {\it ASCA} result.

Results from a {\it Chandra} observation of NGC 5408 X-1 may also
reveal a cool disk in a two-component spectrum (Kaaret et al. 2002).
We speculate that future observations of ULX sources may also reveal
cool accretion disks in some cases.  Future theoretical studies of
anisotropic accretion flows at high accretion rates may be able to
explain such results in terms of stellar-mass black holes.

We wish to thank Albert Kong for his generous communication.  We
acknowledge useful comments from the referee, Jimmy Irwin, and from
Marat Gilfanov, Phil Kaaret, Andrew King, Richard Mushotzky, and,
Zheng Zheng.  J.M.M. acknowledges support from the NSF through its
Astronomy and Astrophysics Fellowship Program.  M.C.M. was supported
in part by NSF grant AST~0098436.  This work is based on observations
obtained with {\it XMM-Newton}, an ESA mission with instruments and
contributions directly funded by ESA Member States and the US (NASA).

\begin{table}[h]
\caption{Spectral Fit Parameters}
\begin{footnotesize}
\begin{center}
\begin{tabular}{llll}
\multicolumn{2}{l}{Model/Parameter} & NGC 1313 X-1 & NGC 1313 X-2\\
\tableline

\multicolumn{2}{l}{power-law} & ~ & ~ \\
\multicolumn{2}{l}{$N_{H}~(10^{21}~{cm}^{-2})$} & $2.0^{+0.2}_{-0.1}$ & $2.7^{+0.3}_{0.2}$ \\
\multicolumn{2}{l}{$\Gamma$} & $1.90\pm 0.05$ & $2.4\pm 0.1$ \\
\multicolumn{2}{l}{Norm. ($10^{-4}$)} & $6.3^{+0.3}_{-0.4}$ & $3.9^{+0.4}_{-0.3}$ \\
\multicolumn{2}{l}{$\chi^{2}/dof$} & 497.7/384 & 177.7/167 \\
\tableline

\multicolumn{2}{l}{MCD} & ~ & ~ \\
\multicolumn{2}{l}{$N_{H}~(10^{21}~{cm}^{-2})$} & $0.41^{+0.09}_{-0.08}$ & $0.6^{+0.2}_{-0.1}$ \\
\multicolumn{2}{l}{$kT$~(keV)} & $1.41\pm 0.06$ & $0.91^{+0.06}_{0.05}$ \\
\multicolumn{2}{l}{Norm. ($10^{-2}$)} & $3.0^{+0.5}_{-0.4}$ & $6.4^{+1.8}_{-1.4}$ \\
\multicolumn{2}{l}{$\chi^{2}/dof$} & 809.5/384 & 265.5/167 \\
\tableline

\multicolumn{2}{l}{Bremsstrahlung} & ~ & ~ \\
\multicolumn{2}{l}{$N_{H}~(10^{21}~{cm}^{-2})$} & $1.27\pm 0.09$ & $1.5\pm 0.02$ \\
\multicolumn{2}{l}{$kT$~(keV)} & $5.9^{+0.6}_{-0.5}$ & $2.6\pm 0.3$ \\
\multicolumn{2}{l}{Norm. ($10^{-4}$)} & $6.4\pm 0.2$ & $3.7\pm 0.3$ \\
\multicolumn{2}{l}{$\chi^{2}/dof$} & 573.3/384 & 205.6/167 \\
\tableline

\multicolumn{2}{l}{Raymond-Smith} & ~ & ~ \\
\multicolumn{2}{l}{$N_{H~(10^{21}~{cm}^{-2}})$} & $1.16\pm 0.08$ & $1.0\pm 0.1$ \\
\multicolumn{2}{l}{$kT$~(keV)} & $5.4\pm 0.3$ & $3.7^{+0.3}_{-0.2}$ \\
\multicolumn{2}{l}{Norm. ($10^{-3}$)} & $1.58\pm 0.05$ & $0.68\pm 0.03$ \\
\multicolumn{2}{l}{$\chi^{2}/dof$} & 749.8/384 & 252.2/167 \\
\tableline

\multicolumn{2}{l}{CompTT} & ~ & ~ \\
\multicolumn{2}{l}{$N_{H}~(10^{21}~{cm}^{-2})$} & $1.1^{+0.3}_{-0.2}$ & $1.7^{+0.7}_{-0.9}$ \\
\multicolumn{2}{l}{$kT_{seed}$~(keV)} & $0.18\pm 0.01$ & $0.15^{+0.05}_{-0.04}$ \\
\multicolumn{2}{l}{$kT_{corona}$~(keV)} & $50^{\dag}$ & $50^{\dag}$ \\
\multicolumn{2}{l}{$\tau$} & $0.80^{+0.07}_{-0.08}$ & $0.37^{+0.07}_{-0.06}$ \\
\multicolumn{2}{l}{Norm. ($10^{-5}$)} & $2.8^{+0.4}_{-0.3}$ & $1.7\pm 0.4$ \\
\multicolumn{2}{l}{$\chi^{2}/dof$} & 431.8/383 & 164.2/166 \\
\tableline

\multicolumn{2}{l}{MCD $+$ power-law} & ~ & ~ \\
\multicolumn{2}{l}{$N_{H}~(10^{21}~{cm}^{-2})$} & $4.4^{+1.2}_{-0.6}$ & $3^{+3}_{-1}$ \\
\multicolumn{2}{l}{$kT$~(keV)} & $0.15^{+0.02}_{-0.04}$ & $0.16^{+0.16}_{-0.04}$ \\
\multicolumn{2}{l}{Norm.} & $1000^{+2100}_{-200}$ & $180^{+80}_{-60}$ \\
\multicolumn{2}{l}{$\Gamma$} & $1.82^{+0.06}_{-0.09}$ & $2.3^{+0.2}_{-0.1}$ \\
\multicolumn{2}{l}{Norm. ($10^{-4}$)} & $6.1^{+0.6}_{-0.9}$  & $3.6^{+1.0}_{-0.8}$ \\
\multicolumn{2}{l}{$\chi^{2}/dof$} & 385.8/382 & 167.1/165 \\
\multicolumn{2}{l}{$F~(10^{-12}~erg~cm^{-2}~s^{-1})^{a}$} & $12^{+10}_{-2}$ & $4.0^{1.1}_{-1.2}$ \\
\multicolumn{2}{l}{$F_{power-law}/F_{total}$} & 0.33  & 0.63 \\
\multicolumn{2}{l}{$L_{0.2-10}~(10^{40}~erg~{s}^{-1})^{b}$} & $2.0^{+1.6}_{-0.3}$ & $0.66^{+0.18}_{-0.20}$ \\
\multicolumn{2}{l}{$L_{0.05-100}~(10^{40}~erg~{s}^{-1})^{b}$} & $3.3^{+4.8}_{-0.3}$ & $1.0^{+0.6}_{-0.1}$ \\

\tableline
\end{tabular}
\vspace*{\baselineskip}~\\ \end{center} \tablecomments{Results of
  fitting simple models to the EPIC MOS spectra of NGC 1313 X-1 and
  NGC 1313 X-2.  The XSPEC model ``phabs''was used to measure the
  equivalent neutral hydrogen column density along the line of
  sight.\\  $\dag$~ The temperature of the up-scattering corona was
  not constrained by fits with the XSPEC model ``CompTT'' and was
  therefore fixed at a reasonable value.\\  $^{a}$ The
  absorption-corrected or ``unabsorbed'' flux.\\ $^{b}$ The luminosity
  in the 0.2--10.0~keV or 0.05--100.0~keV range, assuming a distance of 3.7~Mpc.}
\vspace{-1.0\baselineskip}
\end{footnotesize}
\end{table}

\end{document}